\documentclass[10pt]{revtex4}
\usepackage{hyperref}

\begin{document} 

\title{ Field-dependent symmetries in Friedmann-Robertson-Walker models}
 \author{ Sudhaker Upadhyay} \email {  sudhakerupadhyay@gmail.com; 
 sudhaker@iitk.ac.in} 
 \affiliation{Department of Physics, Indian Institute of Technology Kanpur, Kanpur 208016, India.}

\begin{abstract}
 We consider  effective actions of the cosmological Friedmann-Robertson-Walker (FRW) models   
and  discuss their fermionic rigid BRST invariance.  Further, we demonstrate the finite
field-dependent BRST transformations as a limiting case of continuous field-dependent 
BRST transformations described in terms of continuous parameter $\kappa$.   The
Jacobian under such finite field-dependent BRST transformations is computed explicitly,
which amounts an extra piece  in the effective action within functional integral. We show that
for a particular choice of the parameter the finite field-dependent BRST transformation
maps the generating functional for FRW models from one gauge to another.
\end{abstract}
\maketitle
\section{Introduction}
 The quantum cosmology is a branch of theoretical physics attempting to study the effect of quantum 
mechanics on the formation of the universe, or its early evolution \cite{de,dl}. Despite many attempts, such as the Wheeler-DeWitt equation, and more 
recently loop quantum cosmology, the field remains a rather speculative branch of quantum 
gravity. 
The
cosmological principle  is an axiom that embodies the working assumption or premise that  
  both the spatial homogeneity and the
isotropy of universe   is actually valid
for the very   large scale of the universe rather than the originally stated large scale. The   homogeneous and isotropic spacetime symmetry was originally
studied   by Friedmann, Robertson, and Walker (FRW)  \cite{frw, frw1,frw2,frw3,frw4,frw5} and therefore such universe models are known as the FRW models.
In actual sense, the FRW models are the backbone of modern cosmology describing universe 
because
most of the  works on quantum cosmology are based on the FRW  universe models. However, 
  anisotropic models had also been  studied some time (for instance see \cite{14}).
 Even though almost all
the models of dark energy meet
some difficulties like cosmological constant problems, fine-tuning problems and so on
but they get relevance in FRW spacetime. Therefore, for better
realizations of modern cosmology  a more careful  investigation of  FRW cosmology
is quite demanded.

On the other hand, the realization of gauge symmetry in  FRW models   is well established.
According to standard quantization procedure, the gauge invariant models 
can be quantized correctly by fixing the gauge which removes the 
redundant degrees of freedom in field variables. The well-known path integral procedure to employ gauge-fixing condition at quantum level is known as Faddeev-Popov
trick which involves the so-called Faddeev-Popov ghosts too.
The BRST supersymmetry was introduced in the mid-1970s \cite{brs,brs1} and was quickly understood to justify the introduction of these Faddeev-Popov ghosts and their exclusion from ``physical" asymptotic states when performing  calculations. The BRST symmetry   plays a prominent role in the standard paradigm of fundamental interactions \cite{ht}.

 Although the BRST symmetry has been found for FRW models in particular gauge
  \cite{mon,tp,jj} the generalization of
 BRST symmetry by making the parameter field-dependent, so-called finite field-dependent
 BRST transformation, has not yet been studied.
 The finite field-dependent BRST formulation
 has many applications on gauge field theories \cite{sdj,sdj1,susk,jog,sb1,smm,fs,sud1,rbs,sudhak, rs,sudha,sudha1}.
 So, it is worth analysing such formulation for cosmological models
 describing universe at very   large scale. However, a different kind of 
 field-dependent symmetries in case of non-relativistic fluid model 
 had already been studied \cite{hos}.
This provides us sufficient motivation for present investigation.
In this paper we demonstrate the nilpotent BRST symmetries of 
  the FRW models in various gauges which secures the unitarity of the universe models. Further,
we analyse the aspects of making the parameter of BRST symmetry field-dependent in rather
different way than the finite field-dependent BRST formulation originally advocated in \cite{sdj}.
We found that such revised formulation is simpler than the original one.
Within the analysis we find that
for a particular choice of field-dependent parameter the 
finite field-dependent BRST symmetry connects the generating functional of the model in 
two different gauges.

This paper is outlined as follows.
In Sec. II, we present the FRW models in different gauges with their BRST invariance.
Further,   we analyse the finite   field-dependent BRST symmetry in full generality
 in sec. III. Within Sec. III, we establish the connection between different gauges of FRW models
   using finite   field-dependent BRST symmetry transformation. 
\section{BRST invariant FRW models}
In this section, we discuss the preliminaries of cosmological FRW models describing homogeneous and 
isotropic universe having fermionic rigid BRST invariance. 
So, let us start with the  FRW metric defined in spherical coordinates as follows,
\begin{equation}
ds^2=N^2dt^2+a^2(t)\left(\frac{1}{1-kr^2}dr^2+r^2d\theta^2 +r^2\sin^2\theta  d\phi^2\right),
\end{equation}
where $N$ is the lapse function and $a(t)$ is the unknown potential of the metric that encodes the size at large scales, more formally is the scale factor
of the universe.  Here
both the lapse function and the scale factor depend on time only. However, the values of 
$k=1, -1, 0$ correspond  to a space of positive curvature (closed universe), negative  curvature 
(open universe) and zero curvature (flat universe) respectively. 
Now, we define the classical Lagrangian density of the FRW models  traditionally described in  Arnowitt-Deser-Misner 
(ADM) variables as follows \cite{tp},
\begin{eqnarray}
{ L}_{inv} =  -\frac{1}{2}\frac{a\dot{a}^2}{N}+\frac{k}{2}Na.\label{cs}
\end{eqnarray}
Now, the canonically conjugate momenta corresponding to the lapse function $N$ and the scale factor $a$  
are calculated by,
\begin{eqnarray}
\pi_N &=&0, \label{pr}\\
\pi_a &=&- \frac{a\dot a}{N}.
\end{eqnarray}
The momenta corresponding to $N$ reflect the primary constraint of the theory.
It is now easy to the evaluate the canonical Hamiltonian density (exploiting Legendre transform) \cite{tp}, \begin{eqnarray}
H_c =\pi_a\dot{a}-L_{inv} =-\frac{N\pi_a^2}{2a} -\frac{k}{2}Na.
\end{eqnarray}
Exploiting time conservation of the primary constraint,  we
calculate the   secondary constraint of the theory as follows,
 \begin{equation}
  \frac{\pi_a^2}{2a}+\frac{k}{2}a =0.\label{con}
 \end{equation}
 Since both the constraints (\ref{pr}) and (\ref{con}) are kind of first-class.
 This immediately confirms that the theory of universe embeds gauge invariance.
  The canonical variables transform  under following gauge transformation  \cite{tp}
\begin{equation}
\delta N=-N\dot\eta -\dot N\eta,\ \ \ \delta a=-\dot a\eta,\label{gauge}
\end{equation}
 where   $\eta(t)$ is an infinitesimal   parameter of transformation. 
Now, we follow the standard procedure to quantize the
FRW models. Since before quantizing the theory, it is necessary to impose gauge-fixing condition 
to remove the redundancy in gauge degrees of freedom.
However, the essential requirements for gauge-fixing condition
are as follows: (i) it must fix the gauge completely, i.e., there must not be any residual gauge
freedom, and (ii) using the transformations   it must be
possible to bring any configuration, specified by $N$ and $a$
into one satisfying the gauge condition.
Keeping the above conditions in mind we choose the   
following gauge condition    \cite{tp}:
\begin{equation}
\dot{N} =\frac{d}{dt}f(a),\label{gf}
\end{equation}
where $f(a)$ is an arbitrary function of $a$.
This   gauge condition (\ref{gf}) can be employed  in the theory at quantum level by adding
following  gauge-fixing term in the invariant Lagrangian density (\ref{cs}) \cite{tp}:
\begin{eqnarray}
{L}_{gf}  =   \lambda\left( \dot{N}-\frac{d}{dt}f(a)\right), \label{gaf}
\end{eqnarray}
where $\lambda$ is an auxiliary field.

Now, the determinant corresponding to  the above gauge-fixing term can be compensated in the functional integral by further adding the following ghost term   
 in the effective Lagrangian density:
\begin{eqnarray}
{L}_{gh}  =   \dot{\bar{c}}\left( \dot{N}-\frac{d}{dt}f(a)\right)c +\dot{\bar{c}}N\dot{c},\label{gh}
\end{eqnarray}
where $c$ and  $\bar{c}$ refer the Faddeev-Popov ghost and antighost fields 
respectively. 
Now, the complete extended Lagrangian density
 reads
\begin{equation}
{L}_{ext}={L}_{inv} +{L}_{gf} +{L}_{gh}. \label{tot}
\end{equation}
However, for the different choice of gauge-fixing condition \cite{jj}, 
\begin{equation}
\dot{N} = f(a),\label{gf1}
\end{equation}
the gauge-fixing and ghost terms are demonstrated as follows,
\begin{eqnarray}
{L}'_{gf} &= & \lambda\left( \dot{N}- f(a)\right),\nonumber\\  
{L}'_{gh} &= & \dot{\bar{c}}\dot{N}c +\dot{\bar{c}}N\dot{c}+\bar c \frac{d  }{dt}f(a)
  c.\label{gh1}
\end{eqnarray}
The complete extended Lagrangian density corresponding to the 
gauge condition (\ref{gf1}) is defined by,
\begin{equation}
{L}'_{ext}={L}_{inv} +{L}'_{gf} +{L}'_{gh}. \label{tot1}
\end{equation}
The nilpotent BRST symmetry transformations are constructed by replacing the
parameter $\eta$ of (\ref{gauge}) by ghost field $c$ as follows,
\begin{eqnarray}
s_b N &=& (\dot{N}  c+N\dot c),\nonumber\\
s_b a &=&  \dot{a} c,\nonumber\\
s_b  c&=& 0,\nonumber\\
s_b\bar c&=&-\lambda,\nonumber\\
s_b \lambda &=&0,
\end{eqnarray}
under which both the extended Lagrangian densities ${L}_{ext}$ and ${L}'_{ext}$ are invariant
up to total derivative. 
Since the combination of gauge-fixing and ghost terms for both gauges are
BRST exact and, therefore, we can express these in terms of BRST variation of gauge-fixing fermion
$\Psi$ as follows,
\begin{eqnarray}
{L}_{gf} +{L}_{gh} &= &s_b \Psi= -s_b\left[\bar c\left( \dot{N}-\frac{d}{dt}f(a)\right)\right],\nonumber\\
{L}'_{gf} +{L}'_{gh} &= &s_b \Psi' = -s_b\left[\bar c\left( \dot{N}- f(a)\right)\right],\label{ps}
\end{eqnarray}
where the gauge-fixing fermions have the following expressions $\Psi =-\bar c\left( \dot{N}-\frac{d}{dt}f(a)\right)$ and $\Psi' =-\bar c\left( \dot{N}- f(a)\right)$.

Now, we define the source free generating functional
for FRW models corresponding to gauge conditions (\ref{gf}) and (\ref{gf1}) respectively as
\begin{eqnarray}
Z_1 &=&\int {\cal D}\phi\ e^{iS_{ext}[\phi]},\nonumber\\
Z_2 &=&\int {\cal D}\phi\ e^{iS'_{ext}[\phi]},
\end{eqnarray}
where ${\cal D}\phi$ denotes the generic measure defined in terms of collective field $\phi$
and the effective actions  $S_{ext}$ and $S'_{ext}$ are defined, respectively, by
\begin{eqnarray}
&&S_{ext} =\int d^4x\ L_{ext},\nonumber\\
 && S'_{ext} =\int d^4x\ L'_{ext}.
\end{eqnarray}
Here the Lagrangian densities $L_{ext}$ and $L'_{ext}$ are defined, respectively, in 
(\ref{tot}) and (\ref{tot1}).
\section{Finite field-dependent BRST transformation}
In this section, we demonstrate the methodology of finite field-dependent 
BRST transformation, originally advocated in Ref. \cite{sdj}, in  rather different and elegant 
way.
Then, we discuss its application part.
\subsection{Methodology}
We start discussion by considering the fields $\phi$ as a function of 
  parameter $\kappa: 0\leq \kappa\leq1$ in such manner  that 
the original fields and finitely transformed fields are described by its extremum values.
For instance   $\phi(x, \kappa =0) =\phi (x)$ defines the original fields, however,
   $\phi(x, \kappa =1) =\phi' (x)$ defines the field-dependent BRST transformed fields.
     Now, we define the 
  infinitesimal field-dependent BRST transformation as in \cite{sdj},
  \begin{eqnarray}
  \frac{d\phi(x, \kappa)}{d\kappa} =s_b \phi(x,\kappa) \Theta'[\phi (x,\kappa)].
  \end{eqnarray}
Upon integration the above equation yields  the following continuous field-dependent
    transformation,
  \begin{eqnarray}
  \phi (x,\kappa) =\phi (x,0) + s_b\phi (x, 0)\Theta [\phi(x, \kappa)],
  \end{eqnarray}
which at boundary ($\kappa =1$) leads to the finite field-dependent BRST transformation \cite{sdj},
    \begin{eqnarray}
  \phi' (x) =\phi (x) + s_b\phi (x)\Theta [\phi(x)].
  \end{eqnarray}
Furthermore, we compute the Jacobian of path  integral measure under
 such finite field-dependent BRST transformation by start following the same procedure as discussed in \cite{sdj} as follows,
 \begin{eqnarray}
 {\cal D}\phi (\kappa) = J(\kappa) {\cal D}\phi (\kappa) = J(\kappa +d\kappa) {\cal D}\phi (\kappa +d\kappa),
 \end{eqnarray}
 which further reads
  \begin{eqnarray}
 \frac{J(\kappa)}{J(\kappa +d\kappa) }  = \sum_\phi\pm \frac{{\delta}\phi (\kappa +d\kappa)}{{\delta}\phi (\kappa)},\label{jaco}
 \end{eqnarray}
 where $\pm$ signs are considered suggesting the nature of the fields $\phi$ ($+$ for bosonic
  fields and $-$ for fermionic ones).
 Utilizing the Taylor expansion, the relation (\ref{jaco}) yields \cite{sdj},
 \begin{eqnarray}
 1-\frac{1}{J}\frac{dJ}{d\kappa} d\kappa =1+ d\kappa\int d^4x \sum_\phi\pm s_b\phi(x,\kappa) \frac{\delta\Theta'[\phi(x,\kappa)]}{\delta\phi(x,\kappa)},\label{jac0}
 \end{eqnarray}
Now it is easy to obtain the following expression from the above expression (\ref{jac0}),
 \begin{eqnarray}
 \frac{d\ln J}{d\kappa} =-\int d^4x \sum_\phi\pm s_b\phi(x,\kappa) \frac{\delta\Theta'[\phi(x,\kappa)]}{\delta\phi(x,\kappa)},
 \end{eqnarray}
 Performing further integration, we get the following expression:
  \begin{eqnarray}
  \ln J   &=&-\int_0^1 d\kappa\int d^4x \sum_\phi\pm s_b\phi(x,\kappa) \frac{\delta\Theta'[\phi(x,\kappa)]}{\delta\phi(x,\kappa)},\nonumber\\
  &=&- \left(\int d^4x \sum_\phi\pm s_b\phi(x) \frac{\delta\Theta'[\phi(x)]}{\delta\phi(x)}\right)_{\kappa =1},
 \end{eqnarray}
and consequently we get the exact form of the Jacobian of functional measure as follows:
   \begin{eqnarray}
  J   = {  \exp\left(-\int d^4x \sum_\phi\pm s_b\phi(x) \frac{\delta\Theta'[\phi(x)]}{\delta\phi(x)}\right)}.\label{jac}
 \end{eqnarray}
Hence, with this expression of Jacobian, the generating functional for an effective theory described by an 
effective action $S[\phi]$ changes  under finite field-dependent BRST transformation 
as follows
 \begin{eqnarray}
 \int {\cal D}\phi'\ e^{iS[\phi']} =\int {\cal D}\phi \ e^{iS[\phi ]-\int d^4x \left(\sum_\phi\pm s_b\phi 
 \frac{\delta\Theta'}{\delta\phi }\right)},\label{gen}
 \end{eqnarray}
 where  $\phi'$ refers the transformed fields collectively.
 Therefore, we are able now to draw following conclusion that under whole procedure the effective action of the theory
 gets modified from their original values by an extra piece. However,
in the next subsection, we show  that 
 under such an analysis the  theory does not change
 on physical ground but changes from one convention to another automatically
 which might be useful in computing the physical observable.
 \subsection{An application of finite field-dependent transformation}
 The finite field-dependent BRST transformations for FRW models are constructed as
 \begin{eqnarray}
f_b N &=& (\dot{N}  c+N\dot c)\Theta[\phi],\nonumber\\
f_b a &=&  \dot{a} c\Theta[\phi],\nonumber\\
f_b  c&=& 0,\nonumber\\
f_b\bar c&=&-\lambda\Theta[\phi],\nonumber\\
f_b \lambda &=&0,\label{ff}
\end{eqnarray}
where the  field-dependent BRST parameter is chosen as follows,
\begin{eqnarray}
\Theta [\phi]=\int_0^1 d\kappa\ \Theta'[\phi]= - i\int_0^1 d\kappa\int d^4 x\left[ \bar c \left(\frac{d}{dt}f(a) -f(a)\right)\right].
\end{eqnarray}
Now, corresponding to this $\Theta' [\phi]$,  we calculate the Jacobian for path integral measure 
with the help of formula given in (\ref{jac}) as follows,
   \begin{eqnarray}
  J   &= &{  \exp\left(i\int d^4x \sum_\phi\pm s_b\phi(x) \frac{\delta}{\delta\phi(x)}\left[\bar c \left(\frac{d}{dt}f(a) -f(a)\right)\right]\right)},\nonumber\\
  &=&{  \exp\left( i\int d^4x  \left[\lambda \left( \frac{d  }{dt}f(a) -f(a)\right) 
  +\dot {\bar{c}} \frac{d}{dt}f(a)  c +\bar c  \frac{d}{dt}f(a)  c\right]\right)},\label{exj}
  \end{eqnarray}
 Therefore, under finite field-dependent BRST transformation
 given in (\ref{ff}) the generating functional
 $Z_1$ changes as
   \begin{eqnarray}
 \int {\cal D}\phi'\ e^{iS_{ext}[\phi']} &=&\int J(\phi){\cal D\phi}\ e^{iS_{ext}[\phi]},\nonumber\\
 &=&\int {\cal D}\phi \exp \left( i\int d^4x  \left[ L_{ext} +\lambda \left( \frac{d  }{dt}f(a) -f(a)\right) 
  +\dot {\bar{c}} \frac{d}{dt}f(a)  c +\bar c  \frac{d}{dt}f(a)  c\right] \right), \nonumber\\
  &=& \int {\cal D}\phi \ e^{iS'_{ext}[\phi]} =Z_2.
 \end{eqnarray}
 Here, in intermediate steps, we have utilized the relation (\ref{exj}).
 So, this is nothing but the generating functional of FRW model corresponding to the gauge condition
 (\ref{gf1}). 
 This shows that the field-dependent BRST transformation changes
 generating functional from  one gauge   to  another.
\section{Conclusion}
The well-known models studding   homogeneous and  
isotropic  universe are known as the FRW models.
In modern cosmology, these FRW models have extreme importance as these get relevance 
in most of the dark energy works.
Recently, an interacting and non-interacting two-fluid scenario
for dark energy models  have studied in FRW universe \cite{pra,pra1}.
The models of mass condensation within the
FRW universe lead  to cosmological black holes \cite{fz}.
These cosmological models assume zero cosmological constant that means 
the only force acting is gravity.

We have considered the BRST invariant FRW models describing 
flat, open and closed universe. The BRST symmetries of the models have been
 generalized by making the
transformation parameter field-dependent. 
We have developed the formulation through continuous interpolation
of a parameter $\kappa (0\leq \kappa\leq 1)$ in fields such that
fields at $\kappa =0$ are the original ones, however, at $\kappa =1$ these are the 
 finite field-dependent BRST transformed ones.
The Jacobian for finite field-dependent BRST transformation has been  computed
which depends explicitly on field-dependent parameter. 
 We have found that under finite field-dependent BRST transformation with an appropriate
  choice of field-dependent parameter
 the generating functional for FRW models switches from one gauge to another.
 The present investigation will be useful in development of connection between the two different propagators and also  may be helpful in renormalizing the
 universe models.
 It will be a  step towards the development of full quantum theory of modern cosmology.

\end{document}